\documentclass[preprints,article,accept,moreauthors,pdftex]{mdpi}
\usepackage{trackchanges}
\def\apj{Astrophys. J.}
\def\apjl{Astrophys. J. Lett.}
\def\apjs{Astrophys. J. Suppl.}

\def\prd{Phys. Rev. D}
\def\prl{Phys. Rev. Lett.}

\def\araa{Annu. Rev. Astron. Astrophys}
\def\aap{Astron. Astrophys.}

\def\mnras{Mon. Not. R. Astron. Soc.}

\def\pasj{Publ. Astron. Soc. Japan}
\firstpage{1} 
\pubvolume{1}
\issuenum{1}
\articlenumber{0}
\pubyear{2021}
\copyrightyear{2020}
\datereceived{} 
\dateaccepted{} 
\datepublished{} 
\hreflink{https://doi.org/} 

\Title{Jet-cloud/star interaction as an interpretation of neutrino outburst from the blazar TXS 0506+056}

\TitleCitation{Jet-cloud/star interaction as an interpretation of neutrino outburst from the blazar TXS 0506+056}

\Author{Kai Wang $^{1}$*\orcidA{}, Ruo-Yu Liu $^{2}$, Zhuo Li $^{3}$, Xiang-Yu Wang $^{2}$ and Zi-Gao Dai$^{4,2}$}

\AuthorNames{Firstname Lastname, Firstname Lastname and Firstname Lastname}

\AuthorCitation{Wang, K.; Liu, R.; Li, Z. et al.}

\address{%
$^{1}$ \quad Department of Astronomy, School of Physics, Huazhong University of Science and Technology, Wuhan 430074, China\\
$^{2}$ \quad School of Astronomy and Space Science, Nanjing University, Nanjing, 210093, China\\
$^{3}$ \quad Department of Astronomy, School of Physics, Peking University, Beijing 100871, China\\
$^{4}$ \quad Department of Astronomy, University of Science and Technology of China, Hefei 230026, China\\
}

\corres{Correspondence: kaiwang@hust.edu.cn}

\abstract{A neutrino outburst between September 2014 and March 2015 was discovered from the blazar TXS 0506+056 by an investigation of $9.5$ years of IceCube data, while the blazar was in a quiescent state during the outburst with a gamma-ray flux only about one-fifth of the neutrino flux. In this work, we give a possible interpretation of the abnormal feature by proposing that the neutrino outburst originates from the interaction between a relativistic jet and a dense gas cloud formed via the tidally disrupted envelope of a red giant being blown apart by the impact of the jet. Gamma-ray photons and electron/positron pairs that are produced through the hadronuclear interactions correspondingly will induce electromagnetic cascades and then make the cloud ionized and thermalized. The EM radiation from jet-cloud/star interaction is mainly contributed by the relatively low-energy relativistic protons which propagate in the diffusion regime inside the cloud due to magnetic deflections, whereas the observed high-energy neutrinos ($\gtrsim 100\,\rm TeV$) are produced by the relatively high-energy protons which can keep beamed owing to the weak magnetic deflections, inducing a much higher flux of neutrinos than electromagnetic radiation. The observed low-energy electromagnetic radiations during the neutrino outburst period are almost as same as that in the quiescent state of the source, so it may mainly arise as same as the general quiescent state. As a result, due to the intrusion of a dense cloud, the neutrino outburst can be expected, and in the meantime, the accompanying electromagnetic radiations from hadronic processes will not cause any enhancement in the blazar's electromagnetic flux.}


\begin{document}
	
	\section{Introduction}
	On September 22nd, 2017, a track-like neutrino event IceCube-170922A with energy $\sim 290 \,\rm TeV$ was reported in coincident with a flare of a blazar TXS 0506+056 both spatially and temporally, with a significance at $3\sigma$ level \cite{icecube18a}. Various studies have discussed the possible origin of the event \cite{murase18, liu18, gao18, cerruti18, liao18, magic18, keivani18, zhang18,narek18,righi18,banik19,laha19,padovani19,xue19a,cao20}. If the correlation is true, the discovery indicates effective hadronic processes operate in the jet of TXS~0506+056.
	
	Subsequently, the analysis of historical IceCube data independently shows a $3.5\sigma$ excess of high-energy neutrinos from the same position between September 2014 and March 2015 \cite{icecube18b}. The excess consists of $13\pm5$ events above the expectation from the atmospheric background. Curiously, during this neutrino outburst, the electromagnetic emissions from radio to gamma-ray band of TXS~0506+056 are in the low state. We can infer from such a phenomenon that the jet luminosity is probably not enhanced during the outburst, so the neutrino outburst must be due to the increase of the efficiency of hadronic interactions. In addition, the lack of strong electromagnetic (EM) radiations of this source during the neutrino outburst may favor a hadronuclear origin of the neutrinos over a photohadronic origin \cite{liang18}. On the other hand, the luminosity of gamma rays above $0.1\,\rm GeV$ during the neutrino outburst period is almost the same with its quiescent state, showing only about one-fifth of the luminosity of neutrinos between $32\,\rm TeV$ and $3.6 \, \rm PeV$ \cite{icecube18b, padovani18, liang18}, while the gamma-ray flux generated in the hadronuclear process is supposed to be comparable to the neutrino flux. Although a hard proton spectrum might reproduce such a ratio between gamma-ray flux and neutrino flux, the hard spectrum is not consistent with the neutrino spectrum, unless a spectral break is assumed in the proton spectrum \footnote{even so, the observed gamma-ray spectrum may be still hard to be fitted since the cascade emission induced by the high-energy photon from neutral pion decays may contaminate the gamma-ray emission as well}. Thus, the key and the difficulty to explain the neutrino outburst is to reconcile the measured gamma-ray flux with the neutrino flux in this period.

	
	Motivated by the unusual observations of gamma-rays and neutrinos, some scenarios have been proposed, i.e., the structured blob with an additional compact core~\cite{rodrigues19}, two dissipation blobs located at different distances from the central supermassive black hole (SMBH)~\cite{xue21}, and invoking the produced neutral neutron beam through hadronic processes to suppress subsequent EM cascades contributions~\cite{zhang20}. Suggested by the radio observations, the strong signs of deceleration of jet within the inner region of TXS 0506+056 may be caused by the jet-star interaction \cite{ros20}. In addition, the interaction between AGN jet and red giant (RG) star or dense cloud has been proposed as a possible origin of high-energy radiations from M87 via proton–proton collisions\cite{barkov10,barkov12}. In this work, we study a jet-cloud/star interaction scenario for the neutrino outburst, in which a dense cloud enters the jet and provides additional targets for the hadronuclear interactions (or $pp$ collision) (see the sketch in Fig.~\ref{fig1}). Ref.\cite{liu18} considered clouds in the broad line region (BLR) as targets for hadronuclear interaction to explain the IceCube-170922A, whereas an increased power of the jet is required to meet the brightening of gamma-rays.
	We here consider a scenario that the cloud originates from the tidally disrupted envelope of RG, which moves to the vicinity of the SMBH, being blown apart by the jet \cite{barkov10}. We calculate the hadronuclear interactions between cosmic ray protons and the dense gas in the cloud as well as the EM cascade initiated by the secondary photons and electron/positron pairs. In this paper, we suggest that the EM radiations from jet-cloud/star interaction are mainly contributed by the relatively low-energy relativistic protons which propagate in the diffusion regime due to magnetic deflections inside the cloud, whereas the observed high-energy neutrinos ($\gtrsim 100\,\rm TeV$) are produced by the relatively high-energy protons which can keep beamed owing to the weak magnetic deflections. The high-energy gamma-rays accompanying high-energy neutrinos produced via $pp$ collisions can escape from the cloud but then will be attenuated by extragalactic background light (EBL) and/or cosmic microwave background (CMB). Such an effect can induce that the observed high-energy neutrino flux is much higher than the EM radiations considering the contribution of jet-cloud/star interaction. As a result, the EM signals from jet-cloud/star interaction could be even lower than the quiescent state, predicting non-enhancement of EM radiations, while the neutrino signals could reach the observational level.
	The predicted gamma-ray and neutrino flux will be compared to the measurements in the paper.
	
	
	\section{The requirements of the Cloud}\label{sec2}
	
	The duration of neutrino outburst from the direction of TXS 0506+056 is $t_b={\rm{110}}_{ - 24}^{ + 35}$ days for a Gaussian time window analysis and $t_b={\rm{158}}$ days for a box-shaped time window analysis \cite{icecube18b}. The jet-crossing time of the cloud $t_{\rm jc}$ should be comparable to the duration. Therefore,  by assuming the velocity of the cloud orbiting the SMBH as the Keplerian velocity, the required jet-cloud interaction height from SMBH can be found by $t_{jc}=t_b$, i.e.,
	\begin{equation}
		{z_{\rm jc}} = {\left[ {G{M_{BH}}{{\left( {\frac{{{t_{jc}}}}{{2\theta }}} \right)}^2}} \right]^{1/3}} \simeq 5.5 \times {10^{16}}\theta _{- 1}^{ - 2/3}t_{jc,7}^{2/3}M_{\rm{BH},9}^{1/3}\,{\rm{cm}},
	\end{equation}
	where $\theta$ is the jet semi-opening angle, $M_{\rm{BH},9}=M_{\rm{BH}}/10^9 M_\odot $ (hereafter a notation $x/10^Q = x_{Q}$ is used for a conventional expression in c.g.s. units). One possible origin of the cloud is from an RG, whose external layers are far less gravitationally bound to the stellar core \cite{barkov10}. As a result, the envelope of an RG would suffer significant tidal disruption when the RG passes by the vicinity of an SMBH, and a significant mass $> 10^{30}\,\rm g$ can be unbound from the stellar core \cite{khokhlov93a,khokhlov93b, diener97, ayal00,ivanov03, lodato09}. At the height $z_{jc}$ from SMBH, the RG could lose its outer layers with a radius beyond $R_{\rm RG}^T = {z_{jc}}{({M_{{\rm{RG}}}}/{M_{{\rm{BH}}}})^{1/3}} \simeq 5.5 \times {10^{13}}\theta _{- 1}^{ - 2/3}t_{jc,7}^{2/3}M_{RG, \odot }^{1/3}{\rm{cm}}$ when it penetrates the jet. For a solar-mass RG, the radius can be up to a few hundred of the solar-radius ${R_ \odot }$, so one has $R_{\rm RG} \sim R_{\rm RG}^T$ which corresponds to the ``weak tidal interaction" case in \cite{barkov10}. Under this situation, the blown-apart envelope of RG is still roughly spherical \cite{khokhlov93b}, serving as the required cloud for hadronuclear interactions. The jet-cloud interaction results in a forward shock sweeping through the cloud and heating it. The cloud then would significantly expand to one order of magnitude larger at its sound speed $c_s$ through the mediation between its thermal pressure and the jet pressure. For a jet with a luminosity $L_j$ and Lorentz factor $\Gamma_j$ interacting with a spherical cloud with a radius $r_c$ and a number density $n_c$ at a height $z_{\rm jc}$, by equating the cloud thermal pressure to the jet ram pressure (regardless of the effect of the magnetic field), i.e., $({\Gamma _j} - 1){n_j}{m_p}{c^2} = {n_c}{m_p}v_s^2$, the shock speed $v_s$ can be given by \cite{araudo10}
	\begin{equation}
		{v_s} = {\chi ^{ - 1/2}}c \simeq 3 \times {10^8}n_{c,11}^{ - 1/2}\theta _{ - 1}^{ - 1}z_{jc,17}^{ - 1}L_{j,47}^{1/2}\,{\rm{cm/s}}
	\end{equation}
	as long as $v_s \ll c$, where $\chi=n_c/n_j(\Gamma_j-1)$ is the density ratio of the cloud to the jet. The density of the jet can be estimated to be ${n_j} = {L_j}/\left[ {({\Gamma _j} - 1){m_p}{c^3}\pi R_j^2} \right] \simeq 4 \times {10^5}{L_{j,{\rm{47}}}}{\left( {{\Gamma _j}{\rm{/20}}} \right)^{ - 1}}z_{jc,{\rm{17}}}^{{\rm{ - 2}}}\theta_{-1} ^{-2} \,{\rm{c}}{{\rm{m}}^{{\rm{ - 3}}}} $, where $R_j=\theta z_{jc}$ is radius of transverse section of the jet at height $z_{jc}$. Therefore, the shock crossing time of the cloud is ${t_s} = 2{r_c}/{v_s} \simeq 6 \times {10^6}{r_{c,15}}n_{c,11}^{1/2}{\theta _{- 1}}{z_{jc,17}}L_{j,47}^{ - 1/2}\,{\rm{s}}$.
	\begin{figure}[H]
	\centering
	\includegraphics[width=0.8\columnwidth]{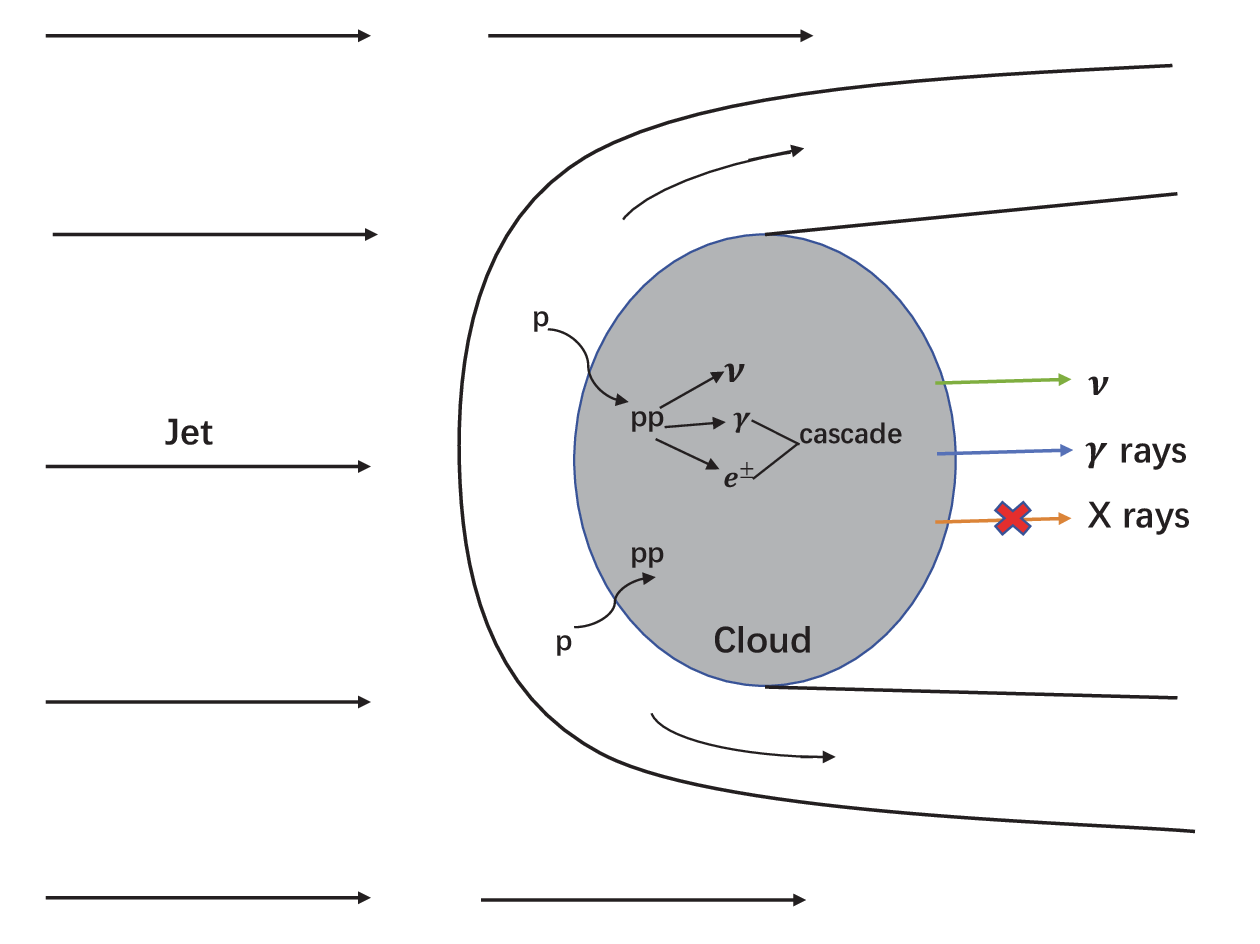}
	\caption{The Sketch (not to scale) of jet-cloud interaction after the intrusion of a cloud into the jet flow. See Section~\ref{sec2} for more details.
	}
	\label{fig1}
    \end{figure}

	Note that the Rayleigh-Taylor (RT) and Kelvin-Helmholtz (KH) instabilities could deform the cloud and mix the materials of the cloud into the jet flow \cite{pittard10}. The timescales of both can be estimated for an instability length $l$ of the perturbation as, ${t_{\rm RT}} \sim \sqrt {l/g}  = \sqrt {4\chi {r_c}l{\rm{/(3}}{{\rm{c}}^2})} $ and ${t_{\rm KH}} \sim \sqrt {l/{g_{rel}}}  = l\sqrt \chi  /c$, where $g = {P_j}\pi r_c^2/{M_c} \sim 3{c^2}/(4\chi {r_c})$ and ${g_{rel}} \sim {c^2}/(\chi l)$ are the acceleration of cloud and the corresponding acceleration of cloud to the jet, and $P_j$ is the jet ram pressure. For the significantly disruptive perturbation $l \sim 2 r_c$, one has $t_{\rm KH}\sim t_{\rm RT} \sim t_s$, which is consistent with some numerical calculations that suggest the cloud can be deformed by the RT and KH instabilities in a timescale $2r_c/v_s$ \cite{gregori00, nakamura06, pittard10}. The fragments of the deformed cloud, nevertheless, could still evolve until all materials of the cloud are melted into the jet flow, and the true lifetime of the cloud (defined as when all of the materials of the cloud are well mixed into the surrounding jet flow) could be more than $6$ times of the timescale of KH instability (see the bottom panel of Figure 15 in \cite{pittard10}). As a result, the true lifetime could be up to $\gtrsim 6 t_{\rm KH}\sim  10^7 \,\mathrm{s}$, which is comparable with the event duration $t_e$. Actually, the effective hadronuclear interactions between the jet flow and the materials of the cloud can take place even if the cloud is melted into the jet flow unless the density of the cloud is significantly decreased. This timescale for effective hadronuclear interactions could be evaluated by the cloud moving from $z_{jc}$ to $2z_{jc}$, so one has $t_{hi} \sim \sqrt{2 z_{jc}/g} \sim 10^7 \rm s$ if neglecting the initial radial speed of the cloud and $g t_{hi} < c$. So both the cloud drag time which is the time needed for accelerating the cloud and the cloud mixing time (or the true lifetime) are several times $t_s$, i.e., $\gtrsim 10^{7}\,$s, for the density ratio of the cloud to the jet considered here.
	
	The hydrodynamic evolution is quite complicated and a detailed discussion is beyond the scope of this work. However, the estimation above for relevant timescales makes it plausible to consider that a dense cloud can exist and provide the targets for effective $pp$ collisions in a timescale consistent with the neutrino outburst duration. Thus, for simplicity, firstly, we consider CRs that accelerated in the jet to penetrate a spherical cloud with a radius of $r_c=5\times 10^{14}\,$cm and a gas density of $n_c=10^{11}\,\rm cm^{-3}$ (see the sketch in Fig.~\ref{fig1}), supplying a high gas column density ${N_H} \simeq {n_c}{r_c} \simeq 5\times {10^{25}} \rm {cm^{-2}}$. The total mass of the cloud can be found by ${M_c} = 4\pi r_c^3{m_p}{n_c}/3 \simeq 8 \times {10^{31}}r_{c,14.7}^3{n_{c,11}}\,{\rm{ g}}$, which can be achieved if the jet's kinetic luminosity is large and the tidal encounter is strong \cite{diener97, ivanov03}. The interaction between the jet flow and the cloud can generate a bow shock and a forward shock crossing the cloud. For the forward shock inside the cloud, the ion plasma frequency is ${\omega _p} = {\left( {4\pi {n_c}{e^2}/{m_p}} \right)^{1/2}} \simeq 4.5 \times {10^8}{n_{c,11}}\,\rm s^{-1}$, while the ion Coulomb collision rate per particle is ${\omega _c} = {n_c}{\sigma _c}{v_p} \simeq 9.8 \times {10^{-3}}{n_{c,11}}{v_{p,9}}{T_{e,8}^{-2}}\,\rm s^{-1}$ by assuming the velocity of a proton is comparable with the shock velocity and the temperature of the proton is same with that of electron (see next Section for details), where $\sigma_c =\pi e^4 /(3kT_e)^{2}$ is the cross section of Coulomb collision. With $\omega_c \ll \omega_p$, the shock is expected to be collisionless in terms of Coulomb interaction. However, accelerations of relativistic protons in the cloud still tend to be prohibited inside the cloud because of the quite large optical depth as discussed in Section~\ref{sec3} (namely, the forward shock is radiation-mediated). Even so, protons can be accelerated to the very high energies in the relativistic jet by some dissipation processes, such as internal collisions between different parts of the jet due to the velocity inhomogeneity, or via the internal-collision-induced magnetic reconnection and turbulence mechanism \cite{zhang11}, or by the bow shock acceleration~\cite{wu22}. The accelerated protons could diffuse into the cloud \cite{taylor61, kaufman90, bultinck10}. By considering the possible advection escape of particles, one can evaluate the advection timescale in the downstream of the bow-shock as ${t_{{\rm{adv}}}} = {r_c}/{v_{{\rm{j,ps}}}} \simeq 6 \times {10^4}{r_{c,14.7}}\,{\rm{s}}$ with the post-bow-shock materials velocity ${v_{{\rm{j,ps}}}} = c/4$. The diffusion timescale is ${t_{{\rm{diff}}}} = {x^2}/2{D_{\rm{B}}} \simeq 1.5 \times {10^5}{({E_p}/1{\rm{TeV}})^{ - 1}}{B_{ - 3}}r_{c,14.7}^2 \,\rm s$ for a Bohm diffusion, where $x = 0.2{r_c}$ and ${D_{\rm{B}}} = {E_p}c/3eB$ \cite{araudo09}. Therefore, the protons with energies $E_p \gtrsim 2.5\, {B_{ - 3}}r_{c,14.7}^2 \,\rm TeV$ could enter the cloud efficiently. Actually, the accelerated protons with relatively low energies can enter the cloud as well, e.g., for the magnetic reconnection acceleration in Ref.~\cite{bosch12}. The advection escape probability could be very small as long as the mean free path of the particle parallel to the magnetic field is much shorter than the characteristic size of the reconnection region and eventually overcome the reconnection layer thickness to enter the cloud \cite{bosch12}. For simplification, we assume the accelerated protons can enter the cloud efficiently.
	These high-energy protons can initiate the subsequent $pp$ collisions, and the characteristic cooling timescale through $pp$ collision in the SMBH rest frame is
	\begin{equation}
		{{t}_{pp}} \simeq 8.4\times 10^{3}\,{n_{c,11}^{-1}}{({\sigma _{pp}}/40{\rm{mb}})^{ - 1}}\,{\rm{s}},
	\end{equation}
	where $\Gamma _j$ is the bulk Lorentz factor of the jet, and $\sigma _{pp}$ is the cross-section of $pp$ collision. The dynamical timescale is 
	\begin{equation}
		{{t}_{dyn}} = {r_c}/c \simeq 1.7\times 10^{4}\,r_{c,14.7}{\,\rm{s}}. 
	\end{equation}
	
	In addition, the relativistic protons may propagate in the cloud by diffusion, especially for relatively low-energy protons. The escape timescale (or residence time in the cloud) can be evaluated by $\tau_{es}=\eta r_c ^2/D_B$, where $D_B =r_g ^2 \omega_g/16$ is the Bohm diffusion coefficient with the gyroradius $r_g= E_p/eB$, the cyclotron frequency $\omega_g=eBc/E_p$ and $\eta \le 1$ is the correction factor that considers the deviation of the actual diffusion from the Bohm diffusion~\cite{kaufman90}. Therefore, one has
		\begin{equation}\label{estime}
			\tau_{es} \simeq 4 \times 10^3 \, \eta_{-1} r_{c,14.7}^2 B_{-3}(E_p/1\,\rm{PeV})^{-1}\,s
		\end{equation}
		for the magnetic field $B=10^{-3}\,\rm G$. Basically, the surface magnetic field strength of most RG stars could be too low (much smaller than $1\,\rm G$) to be detected \cite{lebre14}. However, such a magnetic field strength can be evaluated by the magnetic flux conservation, which indicates that the radial (poloidal) magnetic field component decreases as $B_r \propto r^{-2}$ with the evolution of star radius and the transverse (toroidal) magnetic field component decreases as $B_r \propto r^{-1}$~\cite{spruit01}. For a typical surface magnetic field strength $\sim 1\,\rm G$ of main sequence star, the envelope of RG star can have a magnetic field strength of $\sim 10^{-4} - 10^{-2}\,\rm G$ approximately if the radius of RG star becomes $\sim$100 times larger than that of its main sequence stage. Next, $B=10^{-3}\,\rm G$ in the cloud is adopted for analytical estimations and numerical calculations. The deflection angle of high-energy proton in the magnetic field of cloud can be estimated as $\theta_d\simeq 0.05 \, r_{c,14.7} B_{-3}(E_p/1\,\rm{PeV})^{-1}$ for a propagation distance $\sim r_c$. It is approximately smaller than the jet beaming angle for the high-energy protons with energies beyond $1\,\rm PeV$, i.e., $\theta_d (E_p \gtrsim\,\mathrm{PeV}) \lesssim 1/ {\Gamma} _j=0.05$. As a result, the protons with energies exceeding $\sim 1\,\rm PeV$ propagating inside the cloud, as well as the subsequent generated high-energy neutrinos via $pp$ collisions, can be treated as beamed propagation. Whereas for the protons with energies lower than $\sim 1\,\rm PeV$, they tend to have larger deflection angle and residence time inside the cloud as shown above, inducing a propagation in the diffusion regime. 
		
		As the analysis above, we can see CR protons will lose almost all their energies to secondary particles through $pp$ collision due to the high gas density of the cloud. However, protons with energies $\gtrsim 1\,\rm PeV$ and produced secondary neutrinos can keep beamed while other protons with energies $\lesssim 1\,\rm PeV$ and their secondary neutrinos tend to be more extended in the propagation directions or even isotropic. For simplification, we introduce a correction factor to approximately consider the conversion between the diffuse regime and beamed propagation, namely,
		\begin{equation}\label{diffusion}
			{f_c} = \left\{ \begin{array}{lll}
				1,   & {\theta _d} \le \frac{1}{{{\Gamma _j}}}{\rm{   }}\\
				{\left( {{\Gamma _j}{\theta _d}} \right)^2}, & \frac{1}{{{\Gamma _j}}}<{\theta _d}<2\\
				{\left( {2{\Gamma _j}} \right)^2}, & {\theta _d} \ge 2
			\end{array} \right.
		\end{equation}
		where $\theta_d$ is the defection angle for diverse energy. Our calculations are implemented by treating radiations as beamed and then divided by the correction factor $f_c$.
	\section{Thermalization in the Cloud}
	\label{sec3}
	
	EM cascades will be initiated by high-energy gamma rays and electron/positron pairs generated in $pp$ collision. The electron/positrons generated in the cascades give rise to a strong UV/X-ray emission, and the cloud will be fully ionized by the cascade emission \cite{liu18}. Assuming the mean number density of ionized electrons is the same as that of protons, i.e., $n_{e}=n_c$, the averaged optical depth due to Compton scattering of electrons is
	\begin{equation}
		{\tau _{e\gamma }} = {r_c}{n_{e}}{\sigma _T} \simeq 33\,{r_{c,14.7}}{n_{c,11}}.
	\end{equation}
	Obviously, the cloud is optically thick to the radiations with energies below $\sim\mathrm{MeV}$ \footnote{Above $\sim\mathrm{MeV}$, photons will suffer the Klein-Nishina effect, inducing the optical depths of them will become smaller due to the smaller cross-section in Klein-Nishina regime $\sigma_{KN} \sim \sigma_T (E_{\gamma}/m_e c^2)^{-1}$.}. Due to the shock heating, the proton temperature immediately behind the shock front is $T_p \simeq m_pv_s^2/3k\simeq 4\times 10^8 n_{c,11}^{-1}\,\rm K$. The timescales for Coulombian thermalization through $e-e$ scattering $t_{e-e} \approx 0.1 \,T_{e,8}^{3/2} n_{c,11}^{-1} \,\rm{s}$, $p-p$ scattering $t_{p-p} \approx 3 \, T_{e,8}^{3/2} n_{c,11}^{-1} \,\rm{s}$ and $e-p$ scattering is $t_{ep} \approx 100 \, T_{e,8}^{3/2} n_{c,11}^{-1} \,\rm{s}$ \cite{barkov10}. As we can see, these timescales are quite small and therefore the shocked cloud is thermalized. As a result, we assume proton and electron have the same temperature, i.e., $T_e \simeq T_p$.
	
	The main channel of thermal radiation is the free-free (thermal bremsstrahlung) emission, and 
	the cooling rate of electrons due to the free-free emission can be given by
	\begin{equation}
		{-\left. {\frac{{d{T_e}}}{{dt}}} \right|_{ff}} \simeq {10^{ - 11}}T_e^{1/2}{n_c}\,\rm K/s.
	\end{equation}
	The photon mean energy by free-free emission is $\left\langle \varepsilon  \right\rangle \sim kT_e \sim 10\,\rm keV$ and the total luminosity is ${L_{ff}} \simeq  {10^{44}}T_{e,8}^{1/2}n_{c,11}^2r_{c,14.7}^3\,\rm erg/s$. The concentration of thermal photons can be estimated as
	\begin{equation}
		{n_{ff}} = \frac{{{L_{ff}}}}{{4\pi ck{T_e}r_c^2}} \simeq  {10^{11}}T_{e,8}^{ - 1/2}n_{c,11}^2{r_{c,14.7}}\,\rm {cm^{ - 3}},
	\end{equation}

	Only a fraction of $\sim 1/\tau_{e\gamma}$ of the cascade emission can escape from the cloud without being scattered. Most emitted photons will experience multiple scatterings ($\sim \tau_{e\gamma}^2\sim 1000$ times) inside the cloud.
	A large number of scatterings will lead to the Comptonization of the cascade emission. Energy will be redistributed between photons and electrons and the emergent photon spectrum can be approximated by a Wien distribution at the high-frequency end
	\cite{rybichi79, pozdnyakov83, longair92, ghisellini12, syunyaev80},
	\begin{equation}
		{I_\nu } = \frac{{2h{\nu ^3}}}{{{c^2}}}C{e^{ - h\nu /k{T_e}}}
	\end{equation}
	where the factor $C$ is a constant related to the production rate of the photons and $T_e$ is the temperature of thermal electrons. Such a photon field will in turn influence the EM cascade process so the value of $C$ and $T_e$ is important to the result. The production rate of photons is basically determined by the luminosity of the cascade radiation, which essentially originates from the energy lost by protons in hadronuclear interaction. Given the isotropic-equivalent luminosity of all-flavor neutrino to be $1.2\times 10^{47}\rm \,erg/s$ in the range of 32\,TeV--3.6\,PeV, the beaming corrected luminosity is about $1/\Gamma_j^2$ times smaller, i.e., $7.5\times 10^{43}\,$erg/s with $\Gamma_j=20$. To explain the relatively flat neutrino flux, we need a proton injection with a power-law distribution of a spectral index around $-2$ as suggested by the general Fermi acceleration. Proton energy in each decade is more or less the same for a flat spectral distribution and neutrinos carry about half of the energy lost by protons in $pp$ interactions~\cite{kelner06}, resulting in a bolometric proton luminosity of $\simeq 5\times 10^{44}$erg/s considering the bolometric correction $\ln({E_{p,\rm max}'/m_p c^2})/\ln(3.6\,\mathrm{PeV}/32\,\mathrm{TeV}) \sim 3.5$. Considering that neutrinos carry about half of the energy lost by protons in $pp$ interactions~\cite{kelner06}, the luminosity of cascade which is initiated mainly by absorbed secondary gamma rays and electron/positron pairs ranging from $2-100 \,\rm GeV$ (2\,GeV is corresponding to the minimum energy of the accelerated proton $\sim \Gamma_j m_p c^2 \simeq 20 \,\rm GeV$, $E_{\gamma} \simeq 0.1 E_p$) should be $L_{\rm cas} \simeq 10^{44}\,\rm erg/s$ under the bolometric correction. Assuming the Comptonized photon field reaches a (quasi-)steady state, i.e., the emission rate equal to the energy input rate, we can find the parameter $C$ by
	\begin{equation}
		\pi \int  {4\pi r_c^2{I_\nu }d\nu }  \simeq L_{\rm cas}.
		\label{eq6}
	\end{equation}
	The electron temperature immediately behind the shock front is $m_pv_s^2/3k\simeq 4\times 10^8\,$K (given the high density and high temperature, the time for proton and electron reaching equilibrium via Coulomb collision is extremely short, so we assume proton and electron have the same temperature). The temperature may decrease during the expansion of the cloud. On the other hand, given an average photon energy of $100$eV($\simeq 10^6$K if the distribution is thermal), cascade emissions can heat electrons via Compton scattering if the temperature is too low. Thus, the electron temperature may not drop below $10^6\,$K.
	The electrons in the cloud can be thermalized to be Maxwellian distribution through $e-e$ scattering with the timescales $t_{e-e} \approx 10^{-3}\, T_{e,6}^{3/2} n_{c,10}^{-1}\,\rm{s}$ \cite{barkov10}, the temperature of which can be evaluated by $\frac{3}{2}k{T_e} \simeq \frac{1}{2}{m_e}v_s^2$, i.e., $T_e \approx 2 \times 10^6\,\rm{K} $.

	\section{$pp$ collisions and the cascade emission}
	A fraction of CR protons can enter the cloud and interact with the gas in the cloud. Given a total CR proton luminosity $L_{p,\rm tot}$, and the ratio of the cloud section to the jet section to be $(r_c/R_j)^2=2.5\times 10^{-3}$, the injected CR proton luminosity is $L_p=2.5\times 10^{-3}L_{p,tot}$. Assuming the injected CR protons follow a power-law distribution $\dot{N}_p' = A E_p'^{-s}{\rm exp}(-E'_p/E_{p,\rm max}')$ in the jet comoving frame, we can obtain the normalization factor $A$ by $\int^{E_{p,\rm max}'} A E_p'\dot{N}_p'dE_p'=L_p/\Gamma_j^2$. $s$ is the spectral index and the cutoff energy in the jet comoving frame $E_{p,\rm max}'$ is fixed to be $10^{16}$eV. The spectrum of secondary particles generated in $pp$ collisions is calculated by following the semianalytic method in \cite{kelner06}. The produced high-energy photons and electron/positron pairs will initiate EM cascades in the cloud via the synchrotron radiation, bremsstrahlung, the inverse Compton (IC) scattering, and $\gamma\gamma $ annihilation. The photon number of the Comptonized radiation field can be estimated by
	\begin{equation}
		n_{\rm Comp} \simeq \frac{L_{\rm cas}}{4\pi cr_c^2 \times 3kT_e} \approx 2 \times {10^{11}}{L_{\rm cas,44}}r_{c,14.7}^{ - 2}T_{e,7}^{ - 1}{\rm{c}}{{\rm{m}}^{{\rm{ - 3}}}}.
	\end{equation}
	resulting in an optical depth of $\tau_{\gamma\gamma}\simeq \sigma_{\gamma\gamma}n_{\rm Comp}r_c\simeq 10$ where $\sigma_{\gamma\gamma}\simeq 10^{-25}\rm \, cm^{-2}$ is the peak cross section of the $\gamma\gamma$ annihilation process. The detailed treatment of cascade emission can be found in previous literature, e. g., \cite{wang18}.
	
	The cascade emission in the optical to X-ray band will be scattered via the Compton scattering by thermal electrons, leading to an attenuation of flux in the line of sight by a factor of $(1-e^{-\tau_{e\gamma}})/\tau_{e\gamma}$. In the numerical calculations, a full cross-section of Compton scattering including the Klein-Nishina effect is taken into account \cite{rybichi79}.
	Besides, the very-high-energy (VHE) photons with energies above $100\,\rm GeV$ will be attenuated significantly due to the absorption of EBL by a factor of $e^{-\tau_{\rm EBL}}$. In addition, the different setups of EBL models will not change our results significantly since at such a large luminosity distance $D_L \simeq 1.77 \,\rm Gpc$ for the redshift, i.e., $z=0.3365$. The VHE gamma-rays will be significantly attenuated by the EBL whatever either for optimistic EBL model or lower estimation of EBL model. In the calculation, the employed EBL model here is based on \cite{finke10}.
	
	Note that the VHE gamma-rays can escape from the cloud and the jet flow due to the negligible absorption, especially for the VHE gamma-rays with energies $\gtrsim 100 \,\rm TeV$ which can keep beamed during the propagation. For the $\gamma\gamma$ absorption in the cloud, the required energies of low-energy photons to attenuate the $100\,\rm TeV$ high-energy gamma-rays are around $\sim 0.01\,\rm eV$. The photon number density at $\sim 0.01\,\rm eV$ in the cloud is obviously very small for the adopted electron temperature $10^6-10^8\,\rm K$. In the jet after escaping from the cloud, the required photon energy in the jet comoving frame is $\varepsilon' \sim 0.2\,({\Gamma _j}/20){({E_\gamma }/100{\rm{TeV}})^{ - 1}}{\rm{eV}}$, i.e., $\varepsilon  \simeq {\Gamma _j}\varepsilon ' \simeq 4\,{({\Gamma _j}/20)^2}{({E_\gamma }/100{\rm{TeV}})^{ - 1}}{\rm{eV}} $ in the SMBH frame. $\varepsilon  \simeq 4\,{\rm{eV}}$ is almost the same with the peak energy of the low-energy hump of TXS 0506+056, showing a flux of ${F_\varepsilon } \approx 5 \times {10^{ - 11}}{\rm{erg/c}}{{\rm{m}}^{\rm{2}}}{\rm{/s}}$ \cite{padovani18}. One can consider the radiation region as a jet cone with a semi-opening-angle $1/{\Gamma _j}$ and a radius ${R_f} \sim 2c{\Gamma _j ^2}\Delta {t_{{\mathop{\rm var}} }}/(1 + z) \simeq 8 \times {10^{20}}{\rm{cm}}$, where the time variability is roughly $\Delta {t_{{\mathop{\rm var}} }} \sim 1\,{\rm{ year}}$ in the quiescent state (as shown in Figure 5 in Ref.~\cite{padovani18}). The energy density in the jet comoving frame is
		\begin{equation}
			{{u'}_{\varepsilon '}} \approx \frac{{D_L^2{F_\varepsilon }}}{{\Gamma _j^2 c {R}_f^2}} \simeq 2 \times {10^{-10}}\frac{{{F_\varepsilon }}}{{5 \times {{10}^{ - 11}}{\rm{erg/c}}{{\rm{m}}^{\rm{2}}}{\rm{/s}}}}{\left( {\frac{{{{R}_f}}}{{{8\times {10}^{20}}{\rm{cm}}}}} \right)^{ - 2}}{\rm{erg }}\,{{\rm{cm}}^{{\rm{ - 3}}}},
		\end{equation}
		where $\sigma_{\gamma \gamma } \simeq 10^{-25}\,\rm cm^2$ is the cross section of $\gamma \gamma$ absorption. For the high-energy gamma-rays penetrating this radiation field, the optical depth is (${n'}_{\varepsilon'} \sim {u'}_{\varepsilon '}/ \varepsilon'$) 
		\begin{equation}\label{eqtau}
			\tau  \equiv {{n'}_{\varepsilon '}}{\sigma _{\gamma \gamma }}{({R}_f/\Gamma_j)} \simeq 2.5 \times {10^{ - 3}}\frac{{{F_\varepsilon }}}{{5 \times {{10}^{ - 11}}{\rm{erg/c}}{{\rm{m}}^{\rm{2}}}{\rm{/s}}}}{\left( {\frac{{{{R}_f}}}{{{8\times {10}^{20}}{\rm{cm}}}}} \right)^{ - 1}},			
		\end{equation}
		which is too small to absorb the high-energy gamma-rays. Note that the above estimate of the optical depth may be optimistic for a approximate time variability $\sim 1\,\rm year$ for the quiesent state. The variability timescale may be smaller e.g., $\sim 1\,\rm week$ in Ref.~\cite{morokuma21} for the follow-up observations for IceCube-170922A, resulting in a smaller emission radius. If a smaller variability timescale is involved, a larger optical depth will be expected as described in Eq.~\ref{eqtau}, e.g., $\tau \simeq 0.1$ for a variability timescale of $1\,\rm week$. In addition to $\gamma \gamma$ absorption, for the electron scattering, these high-energy gamma-rays can escape from the cloud as well due to the suppressed scattering cross-section in the Klein-Nishina regime. However, these high-energy gamma-rays can be absorbed significantly by EBL and CMB for a redshift ~0.3365, resulting in electromagnetic cascades in the intergalactic medium. The deflection of electrons by the intergalactic magnetic field is expected to spread out the cascade emission and consequently contribute little to the observed flux. A strong intergalactic magnetic field $\gtrsim 3 \times 10^{-16}\,\rm G$ has been suggested due to the non-detection of the extended GeV halo around blazars, which deflects the pairs out of the line of sight prior to their IC emission \cite{ackermann18} \footnote{In this situation, the accompanying degree-scale IC halos may still exsit, but the absence of this feature in the observations as claimed in Ref~\cite{broderick18} may imply other dominant physical processes that preempt the creation of halos, e.g., the presence of beam-plasma instabilities in the intergalactic medium}. Therefore, we neglect the contribution of intergalactic cascades.
	
	\begin{figure}[H]
		\centering
		\includegraphics[width=1.0\columnwidth]{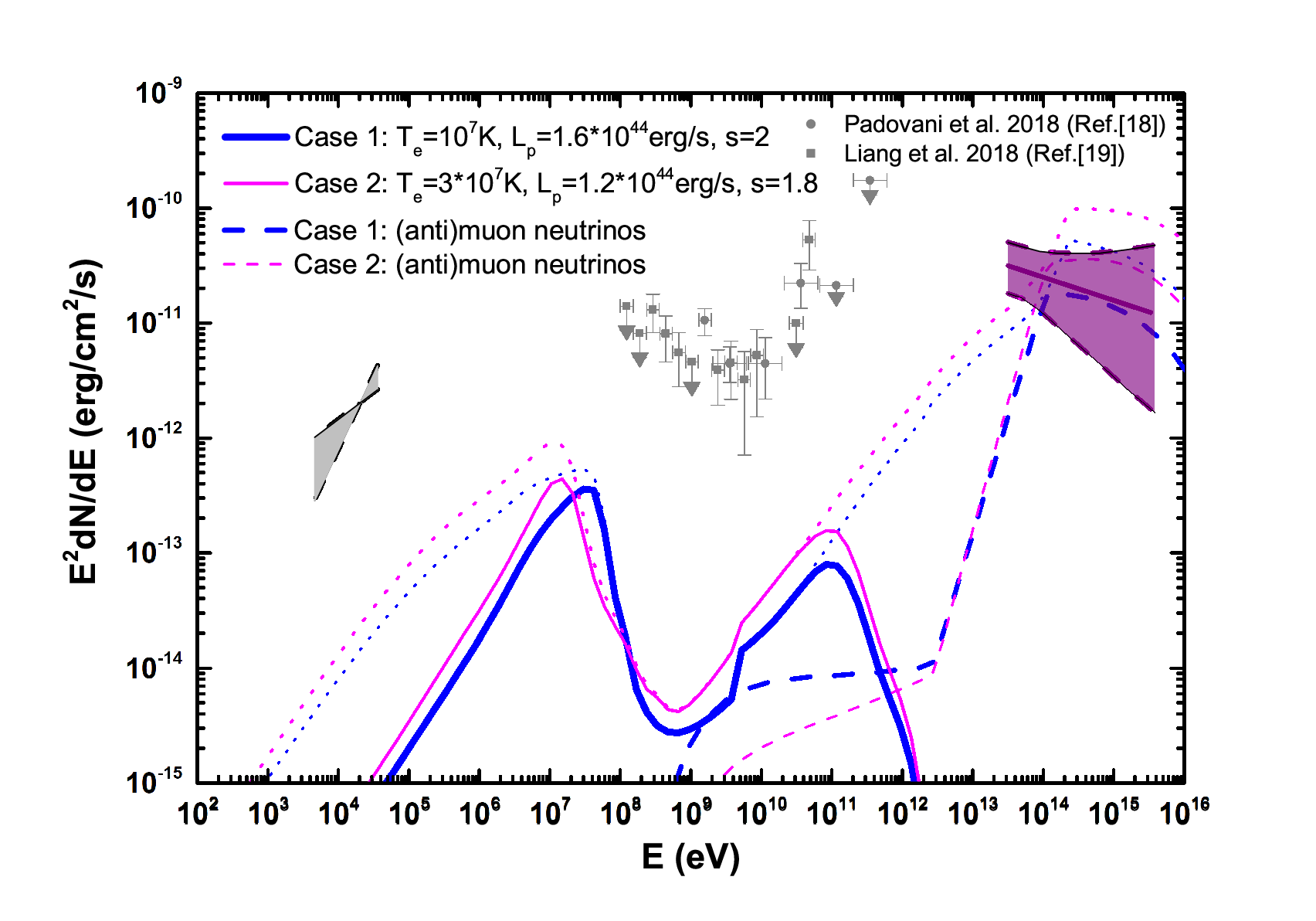}
		\caption{The predicted photon flux (solid curves) produced by secondary electromagnetic particles generated in $pp$ collision and the corresponding neutrinos flux (dashed curves). The blue curves are for an electron temperature of $T_e=10^7\,$K in the cloud with spectral index $s=2$ of the injected CR protons, while the magenta ones are for $T_e=3\times 10^7\,$K and $s=1.8$. The purple-shaded region represents the uncertainty of observed neutrino flux and the gray region represents the X-ray flux in archival data. The gray circular data points are gamma-ray flux during the neutrino outburst analyzed by Ref.~\cite{padovani18} and the square ones are analyzed by Ref.~\cite{liang18}, which are almost the same as the observations in the quiescent state. The blue and magenta dotted curves represent the photon flux before the absorption by EBL and the Compton scattering by thermal electrons in the cloud. In both cases, except adopted parameter values shown in the figure, other parameters are the same, i.e., $L_p=1.4\times 10^{44} \,\rm erg/s$, $E'_{p,\max}=10^{16}\,\rm eV$,  $B=10^{-3} \,\rm G$, $r_c=5\times 10^{14} \,\rm cm$, $n_c=10^{11} \,\rm cm^{-3}$, $\Gamma_j=20$ and the redshift $z=0.3365$.
		}
		\label{fig2}
	\end{figure}
	
	Our results are shown in Fig.~\ref{fig2}. The blue solid curve represents the predicted photon flux with the cloud temperature of $T_e=10^7\,$K and an injection spectral index of $s=2$ for CR protons. There is a dip in the spectrum around 0.3\,GeV due to the absorption by the Comptonized photon field.
	The absorption leads to a hard spectrum above the 0.3\,GeV. The blue dashed curve shows the muon and anti-muon neutrino flux assuming a flavor ratio of $1:1:1$ after oscillation. The blue dotted curve shows the photon flux without Compton scattering and EBL absorption for reference. The results with $T_e=3\times 10^7\,$K and $p=1.8$ are shown with magenta curves. Comparing the magenta curves with blue curves,
	the dip is shallower with a higher temperature due to the photon number density being smaller given the same energy density. We also plot the Fermi-LAT data analyzed by Ref.~\cite{padovani18} and by Ref.~\cite{liang18} in the figure. The X-ray and gamma-ray emission (gray data in Fig.~\ref{fig2}) are almost the same as the observations in the quiescent state, presenting a non-enhancement behavior. The EM radiations from jet-cloud interaction present a quite low flux since they are mainly generated by the relatively low-energy CR protons ($\lesssim 1 \,\rm PeV$) that become more diffused or even isotropic due to the magnetic deflection inside the cloud. However, the high-energy neutrinos ranging from tens of TeV to multi-PeV produced by the relatively high-energy CR protons ($\gtrsim 1\,\rm PeV$) can keep beamed due to very weak magnetic deflections. Therefore, neutrinos have a much higher flux than the EM radiations only considering contributions from jet-cloud interaction. The very fast increase of neutrino flux between multi-TeV to $\sim 100\,\rm TeV$ shown in Fig.~\ref{fig2} is due to the correction factor of Eq.~\ref{diffusion} considering the transformation from diffusion regime to beamed propagation. In addition, the high-energy gamma rays with energies larger than 100 GeV are attenuated significantly by the EBL. Eventually, when the neutrino signals for both cases in Fig.~\ref{fig2} can reach the observational level, the EM radiations from jet-cloud interaction can be much lower than the quiescent state, predicting non-enhancement of observed EM radiations. We speculate that the observed EM emission arises from its general quiescent state, while the neutrino outburst in this time duration is produced by the sudden intrusion of a dense cloud. To match the observed high-energy neutrino flux, the required luminosity of CR protons injected into the cloud is $L_p\simeq 10^{44}\,$erg/s, which translates to a total luminosity of CR proton in the jet to be $(5-6)\times 10^{46}\,$erg/s. It is smaller than the Eddington luminosity $L_{Edd}\simeq 1.3 \times 10^{47} M_{\rm{BH}, 9 }\,\rm erg/s$ of a $10^9M_\odot$ SMBH, in accordance with the fact that the source is in the quiescent state during the neutrino outburst. 
	
	Even though the adopted luminosity in our model is smaller than the Eddington luminosity, it has reached $\sim 50\%$ of the Eddington luminosity for a $10^9M_\odot$ SMBH. However, BL Lac objects are generally believed to have a low Eddington ratio. For the leptonic models of blazar radiations, a sub-Eddington jet power is required due to the high radiation efficiencies of electrons. A super-Eddington jet power is needed when hadronic models with low efficiencies are involved \cite{zdziarski15}. However, in our model, the efficiency of the hadronic model is close to $100\%$ owing to the dense gas target and then a sub-Eddington ratio is invoked. Besides, the Eddington luminosity is achieved by the balance of radiation pressure and gravity, which is not a strict limit on the luminosity of a black hole since some mechanisms, e.g., the Blandford-Znajek mechanism \cite{blandford77} which extracts the spin power of SMBH and the super-critical accretion of SMBH \cite{beloborodov98,sadowski15}, could break through the Eddington luminosity. In addition, Ref~\cite{padovani19} claims the blazar TXS 0506+56 is masquerading BL Lac, namely, intrinsically a flat-spectrum radio quasar with hidden broad lines, implying a possible high Eddington ratio.
	

We also tried other model parameters. For the estimated electron temperature ranging $\sim 10^6- 10^{8}\,\mathrm{K}$, the cascade emissions will not exceed the quiescent state since the significant isotropization of low-energy EM radiations when the high-energy neutrino flux is consistent with observations. The cutoff energy of protons in the jet comoving frame $E_{p,\rm max}'$ is fixed to be $10^{16}$eV, namely, $E_{p,\rm max} = \Gamma_j E_{p,\rm max}'= 200\,\rm PeV$ in the SMBH rest frame, which can roughly generate neutrino with the highest energy $E_{\nu,\max} \simeq 0.05 E_{p,\rm max} \simeq 10 \,\rm PeV$. Thus, $E_{p,\rm max}'=10^{16}$ can satisfy the requirement of the observational upper limit of neutrino energy, i.e., $7.5 \,\rm PeV$. For a larger $E_{p,\rm max}'$, differences in results would be tiny since the normalization factor $A$ will change slightly for a flat CR spectrum. For a smaller $E_{p,\rm max}'$, the produced neutrino can not reach the upper limit of energy range provided by IceCube. The adopted value of the magnetic field is prominent since it determines the characteristic energy of conversion from the diffusion regime to the beamed propagation, which is important to reduce EM radiations. For a larger magnetic field, higher energy protons would propagate in the diffusion regime and make the observed neutrino number lower and vice versa. Basically, one has the magnetic field $B \lesssim {10^{ - 3}}r_{c,14.7}^{ - 1}{\left( {\frac{{{\Gamma _j}}}{{20}}} \right)^{ - 1}}\frac{{{E_p}}}{{1\,\rm PeV}}\,\rm G$ to keep protons with energy $E_p$ beamed. Our adopted magnetic field strength is located at the reasonable range of magnetic field of RG envelope as mentioned in Section~\ref{sec2}.

\begin{figure}[H]
\centering
\includegraphics[width=1.0\columnwidth]{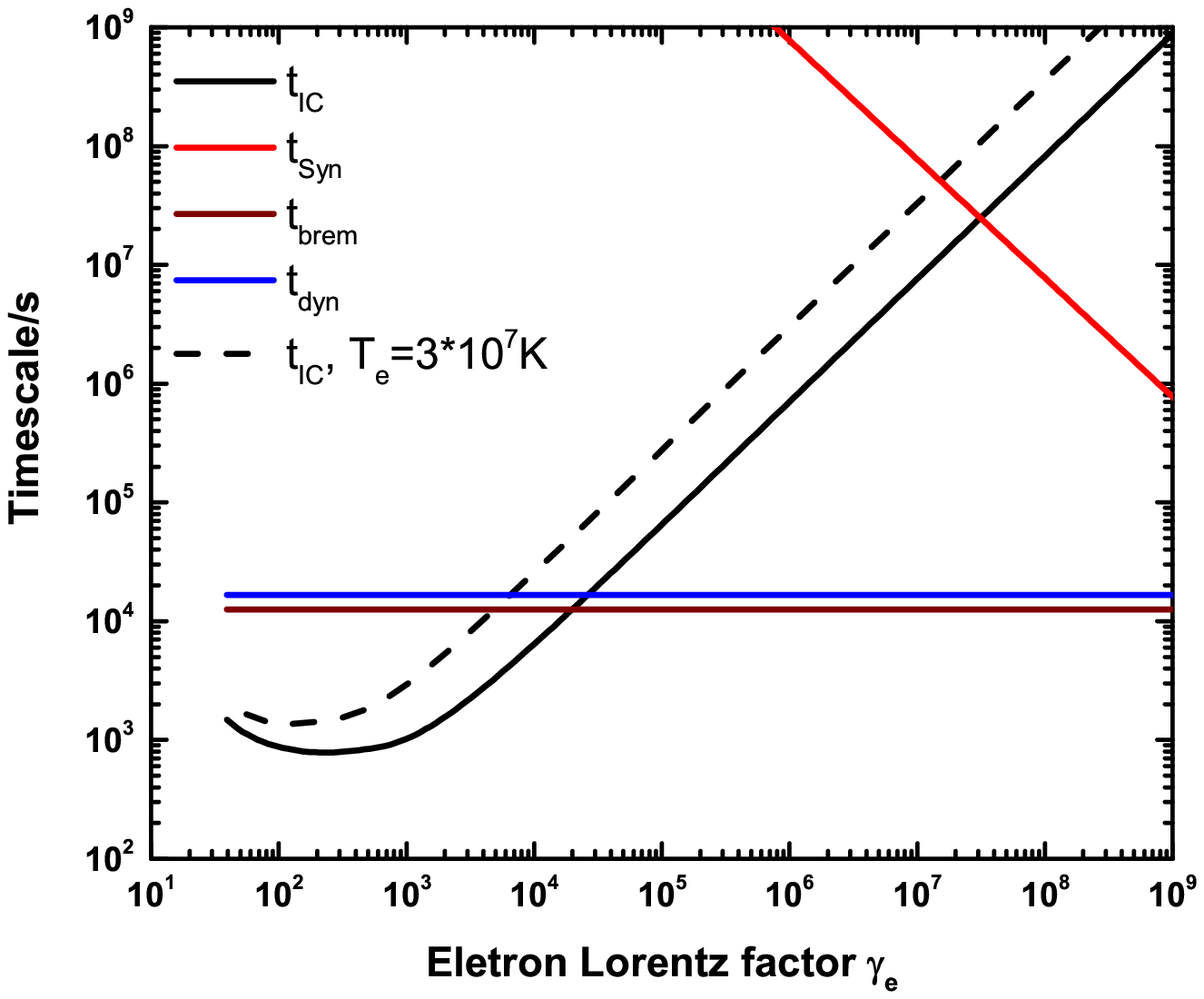}
\caption{The synchrotron, IC, bremsstrahlung and dynamical timescales of electrons in the SMBH rest frame. the same parameters are adopted as in Fig.~\ref{fig2}.
}
\label{fig3}
\end{figure}

In addition, for secondary electrons in the cascade emission, the electron timescales of synchrotron, IC, bremsstrahlung and dynamical evolution are presented in Fig~\ref{fig3}, which indicates that the IC and bremsstrahlung processes dominate at the lower and higher electron energy for the adopted parameter values, respectively. Note that the Comptonzied radiation field also provides target photons for the photomeson process. However, given a $\gamma\gamma$ opacity of $\sim 10$, the efficiency of the photomeson interaction is of the order of $0.01$. Although the radiation field can permeate the entire jet section and interact with all the CR protons in the jet which is $(R_j/r_c)^2$ times more than CR protons that enter the cloud, this factor is canceled by the fact that the photon field density also decreases by a factor of $(r_c/R_j)^2$ at the scale of the entire jet section. Thus, we can neglect the photomeson process.

\section{Discussion and Conclusion}
During the period of neutrino outburst, the EM emissions including X-rays and gamma-rays of TXS~0506+056 are in the low state. This leads us to consider a jet-cloud/star interaction scenario in which the outburst is due to the increase of efficiency of hadronuclear interaction rather than the increase of jet power. The observational gamma-ray flux is much lower than the high-energy neutrino flux. It violates the prediction of the hadronic process which generally generates the comparable gamma-ray and neutrino flux. The X-ray to gamma-ray flux also put an upper limit on the emission of the EM cascade initiated by hadronic interactions. Bearing these requirements in mind, we showed the jet-cloud/star interaction model could successfully explain the neutrino outburst. The intrusion of a dense cloud into the jet flow serves as the target for hadronic processes of high-energy protons, generating high-energy neutrinos and gamma-rays effectively. The EM radiation from the jet-cloud interaction is expected to be spread out since they are mainly contributed by the relatively low-energy protons that propagate in the diffusion regime inside the cloud, reducing the expected EM flux to the level below the observational upper limit. However, the high-energy neutrinos in the IceCube energy band are produced by the relatively high-energy protons that can keep beamed due to the weak magnetic deflections. As a result, for our model, during the outburst of high-energy neutrinos, the accompanying EM radiations from hadronic processes will not cause any enhancement in the blazar's EM flux.


The magnetic field strength is crucial to determine the characteristic energy of conversion from the diffusion regime to the beamed propagation in our model. A smaller magnetic field is important to make the expected high-energy neutrino flux much higher than the accompanying EM flux, inducing a consistent result to the observations. However, a too large magnetic field can lead to the diffusion of produced high-energy neutrinos as well, causing our model to be invalid. Moreover, our model predicts a dip in the gamma-ray spectrum which is due to the $\gamma\gamma$ annihilation by the Comptonzied cascade radiation. It results in a hardening in the spectrum beyond the dip energy which is seen in this event. On the other hand, we also expect a relatively high flux at $\sim10\,\rm MeV$, which is in the detectable energy range of e-ASTROGAM. These gamma-ray features may not be detectable either for a Blazar or for other distant sources due to the relatively low flux. They could be overshot by EM emissions from the blazar or too weak to be detected for a distant source. However, they may be detectable for a nearby non-Blazar AGN, like M87 and Cen A \cite{barkov12}.
The temperature of the cloud is crucial to the position and depth of the dip. It is estimated to be $10^6-10^8\,$K in this work, leading to the energy of the dip in the range of $0.01-10$GeV. The temperature may be measured from the spectrum of Comptonized radiation. However, the Comptonized radiation is isotropic, resulting in a flux of the order of only $10^{-13}\rm \, erg/cm^2s$ in X-ray for TXS 0506+056, and is outshone by the nonthermal emission of the jet. However, it may be observable from nearby misaligned AGNs if jet-star/cloud interactions happen, probably accompanying a brightening of the TeV emission due to the negligible absorption of EBL. This may also serve as a test for our model in the future.

In addition to the blazar TXS 0506+056, the prediction of a much higher neutrino flux than the EM signal in our model could be responsible for other "orphan" neutrino flares, e.g., IceCube-200107A with the blazar 4FGL J0955.1+3551 \cite{giommi20}. The inner regions of galaxies usually contain a large amount of gas, dust, and stars \cite{burbidge70}. Assuming that the active galaxies are similar to our Milky Way, in the inner region of a galaxy, the stellar mass density is $\sim 2.6 \times {10^7}\,{M_ \odot }\rm p{c^{ - 3}}$ within a distance of 0.01 pc from the central black hole~\cite{schodel18}. Thus, in one galaxy, the number of RGs within 0.01 pc can be roughly estimated as $\sim 0.26$ under the assumption that the mass of a star in this region is one solar mass, $\sim 1\%$ of stars are RGs \cite{young77}. For the known $\sim 3500$ Fermi-LAT blazars \cite{ajello20} and a typical jet opening angle $\sim 0.1 $, the blazar jet-RG interaction number in the universe is roughly $\sim 2$. Note that this estimated interaction number is very crude since the active galaxies may be very different from Milky Way. Even the jet-RG interaction can occur, diverse situations, e.g., the magnetic field, the mass of RG, the position of RG intrusion, and the jet power, etc., could make EM and neutrino radiations very different.
\vspace{6pt} 



\authorcontributions{Conceptualization, K.W. and R.L.; investigation, K.W. and R.L.; writing—original draft preparation, K.W.; writing—review and editing, R.L., Z.L., X.W. and Z.D.; supervision, Z.L., X.W. and Z.D.. All authors have read and agreed to the published version of the manuscript.}

\funding{This work is supported by the National Basic Research Program of China (973 Program, Grant No. 2014CB845800), the National Key Research and Development Program of China (Grant No. 2017YFA0402600 and 2018YFA0404200), the National Natural Science Foundation of China (Grant No. 11773003, 11573014, 11625312, 11851304 and 12003007) and the Fundamental Research Funds for the Certral Universities (No. 2020kfyXJJS039).}

\institutionalreview{Not applicable.}

\informedconsent{Not applicable.}

\dataavailability{Not applicable.} 

\acknowledgments{We would like to thank the anonymous referees for very helpful comments and useful suggestions.}

\conflictsofinterest{The author declares no conflict of interest.}


\end{paracol}
\reftitle{References}

\end{document}